\documentclass{emulateapj}

\usepackage{graphicx}
\usepackage{amsmath}
\usepackage{amssymb}
\usepackage{latexsym}
\usepackage{epsfig}

\shorttitle{Diffuse X-ray properties of ZENS galaxy groups}
\shortauthors{Miniati et al.}

\begin{document}
\title{The X-Ray Zurich Environmental Study (X-ZENS). II. X-ray Observations of the Diffuse Intragroup Medium in Galaxy Groups}
\author{
Francesco Miniati\altaffilmark{1},
Alexis Finoguenov\altaffilmark{2},
John D. Silverman\altaffilmark{3},
Marcella Carollo\altaffilmark{1},
Anna Cibinel\altaffilmark{4},
Simon J. Lilly\altaffilmark{1},
Kevin Schawinski\altaffilmark{1}
}
\altaffiltext{1}{Physics Department, Wolfgang-Pauli-Strasse 27,
ETH-Z\"urich, CH-8093, Z\"urich, Switzerland; fm@phys.ethz.ch}
\altaffiltext{2}{Department of Physics, Gustaf H\"allstr\"omin katu 2a,
FI-00014, University of Helsinki, Finland}
\altaffiltext{3}{Institute for the Physics and Mathematics of the Universe (IPMU), University of Tokyo, Kashiwanoha 5-1-5, Kashiwa-shi, Chiba 277-8568, Japan}
\altaffiltext{4}{CEA Saclay, DSM/Irfu/S\'ervice d'Astrophysique, Orme des Merisiers, F-91191 Gif-sur-Yvette Cedex, France}


\begin{abstract}
We  present the results of a pilot XMM-$Newton$
  and $Chandra$ program aimed at studying  the diffuse intragroup medium
  (DIM) of optically-selected nearby groups from the Zurinch ENvironmental Study (ZENS) catalog. The groups  are in a
  narrow mass range about $10^{13}M_\odot$, a mass scale at which the interplay between the DIM and the group member galaxies is still largely  unprobed.  X-ray emission from the DIM is detected in the energy band 0.5--2 keV with flux $\le 10^{-14}$ erg cm$^{-1}$ s$^{-1}$, which is one order of magnitude fainter than for
typical ROSAT groups (RASS). For many groups we set upper limits to the X-ray luminosity, indicating that  the detections are likely probing the upper envelope of the X-ray emitting groups. We find evidence for our optically selected groups to be under-luminous with respect to predictions from X-ray scaling relations.
X-ray mass determinations are in best agreement with those based on the member galaxies bulge luminosity, followed by their total optical luminosity and velocity dispersion. We
measure a stellar mass fraction with a median value of about 1$\%$, with a
  contribution from the most massive galaxies between 30 to 50 \%. 
 Optical and X-ray data give often complementary answers concerning the dynamical state of the groups,
 and are essential for a complete picture of the system. 
Extending this pilot program  to a larger sample of groups is mandatory to unveil any imprint of  interaction between member galaxies and DIM  in halo potentials of key importance for environmentally-driven galactic evolution. 
\end{abstract}

\section{Introduction} \label{intro:sec}
Galaxy groups in the mass range $10^{13}-10^{14}M_\odot$
are key structures in the cosmic fabric.
Containing typically between a few to a few tens of member
galaxies~\citep{Mulchaey00,Mulchaey03}, they 
are significantly more numerous than galaxy
clusters, trace the filamentary components of the large scale
structure \citep[e.g.][]{Eke04} and are likely to contain a large fraction of the
unaccounted cosmic baryons~\citep{FukugitaHoganPeebles98}. 
Groups have historically been detected in the X-ray emission,
typically extending on scales of a few hundred kpc, revealing a gas in
a state of diffuse, ionized and metal rich plasma with temperature
around one keV.  The plasma emission mechanism is a combination of
thermal bremsstrahlung and line emission, the latter more prominent
in groups than in clusters of galaxies due to the lower plasma temperature. The
detection of hot plasma bears witness to the fact that groups are not
simply associations of galaxies by way of projection effects, but real
physical systems, gravitationally bound and having undergone some
degree of virialization.

Groups host about 50 to 70\% of today's $L*$ galaxies~\citep{Eke05}, thus
providing the environment most commonly experienced by the latter.
The role and impact of group environment on galaxies is therefore
subject of detailed studies in both optical and X-ray surveys
 \citep[e.g.][ among the others]{Rasmussen06,Weinmann06,Poggianti08,McGee11,Knobel12,Giodini12,Carollo13}.
Theoretical models indicate that groups
can be instrumental in triggering processes that accelerate galaxy
evolution.  For example, the in-spiral timescale of dynamical friction
varies in proportion to $\sigma^3$/$\rho$, where $\sigma$ and $\rho$
are the halo velocity dispersion and density, respectively.  Thus,
compared to the field and galaxy clusters, galaxy tidal interactions
due to close encounters (harassment) and mergers take place on a
cosmologically short timescale in groups, characterized by significant
over-density and relatively low velocity
dispersion~\citep{Barnes90}.  Such galaxy-galaxy interactions induce
dynamical instabilities~\citep{Mayer01}, plausibly allow for the
fueling of central super-massive black holes
~\citep[SMBHs,][]{BarnesHernquist91,Barnes02,Mihos03} and are likely to play a major role in
 the evolution of massive galaxies within the group environment \citep{McIntosh2008,Feldmann10,Robotham13}.
In addition, gas dynamical interactions with the diffuse intra-group medium (DIM),
such as ram-pressure stripping of cold gas in the stellar disk and/or
strangulation of the baryonic supply from the halo reservoir, can
severely reduce the gas available for star formation and feedback
process even in relatively low density
environment~\citep{Kawata08,McCarthy08}, possibly leading to  the observed satellite quenching in groups
\citep[e.g.,][]{van_den_Bosch08,Skibba09,Peng12,Wetzel13,Carollo14,Tal14}.

The group mass sets an important scale, one at which the processes of
large scale structure formation and galaxy formation meet. In fact,
the thermodynamic properties of the DIM deviate substantially from
scaling relations obeyed by massive clusters, an indication that
processes other than gravity and (magneto-) hydrodynamics affect in an
important way the gas energetics at this mass scale.  Particular
attention has been devoted to the study of gas entropy, because it is
conserved during adiabatic processes and changes only according to the
net amount of thermal energy either absorbed or released by the gas.
Gas entropy in groups appears in clear excess with respect to the
expected value from the scaling relations obeyed by clusters,
~\citep{PonmanCannonNavarro99,Lloyd-DavisPonmanCannon00,FinoguenovArnaudDavid01,PonmanSandersonFinoguenov03,Madhavi05,Pratt10},
a fact ascribed, at least in part, to energetic feedback from galaxy
formation
processes~\citep[e.g.,][]{Voitetal02,CavaliereMenciTozzi98,Borganietal05,McCarthy10,Giodini10}.
In fact, due to the group's shallow potential well and corresponding
low virial temperature, the energy released by past star formation and
AGN activity leaves a distinct footprint on the thermodynamic
properties of the DIM.  In addition, the DIM appears highly enriched
with significant amounts of metals, most likely transported there from
the parent galaxy through stellar
winds~\citep{RasmussenPonman09,McCarthy10}.  Winds associated with
metal transition lines observed at redshift $z\sim1$ appear to be
strongly magnetized~\citep{Bernetetal08,Bernetetal10,Bernetetal12}, so
the DIM is likely to be significantly magnetized as well, in addition
to being metal rich.

The above interplay between galaxies and DIM is important because it
affects the observational properties of X-ray groups as well as those
of galaxies.  Its study is important in shedding light on the origin
of the observed physical properties of the DIM, including amongst
others, the radial profile of thermodynamic quantities such as
entropy, temperature, and their dependence with virial mass (or any of
its proxies).  Likewise, the feedback processes described above, are
believed to play a crucial role in galaxy evolution.  In particular
current models suggest that the growth of galaxies is intimately
related to the growth of their black holes, and that AGN activity is
key in preventing excessive cooling in massive galaxy
halos~\citep{Bower06,Croton06,Bower08}.

From an observational point of view, the study of the interplay
between galaxies and the DIM remains incomplete for two basic reasons.
First, it is now well known that groups are a heterogeneous class,
spanning a wide range in dynamical state~\citep{Mulchaey00} and X-ray
emission~\citep{EkcmillerHudsonReiprich12}. Because X-ray observations
provides the most straightforward  way for identifying groups, X-ray selected
group samples are traditionally the best
studied~\cite[e.g.,][]{HeldsonPonman00,FinoguenovArnaudDavid01,Madhavi05,OsmondPonman04,RasmussenPonman07},
and  have delivered many important results.
However, dynamically relaxed groups tend to be more X-ray bright, which 
raises an important and well appreciated issue concerning the
{\it representative} character of X-ray selected group samples. At cluster scales, several studies have indeed shown that the X-ray properties of 
optically-selected galaxy clusters differ substantially from those of the
X-ray-selected structures~\citep{Donahue02,Gilbank04,Lubin04,Popesso07}.
With the advent of large spectroscopic surveys, e.g., the
SDSS~\citep{York00} and the 2dFGRS~\citep{Colless01}, it has
become possible to identify also group-sized structures spectroscopically, i.e. through the
identification of their member galaxies.  Pioneering studies on a few systems have demonstrated the existence, also at the mass scales of groups, of less relaxed structures with low X-ray emissivity~\citep{Mulchaey00,Rasmussen06} that were missed in
shallow X-ray surveys (RASS).

As for the second reason for the above incompleteness, studying galaxy
groups up to now has been limited by the lack of accurate information
about the properties of the galaxy-members together with the X-ray
data.  It is now clear, however, that in view of the rich variety of
properties of galaxies observed at fixed group mass scale as a
function of galaxy mass and of rank within the groups as central or
satellite~\citep{Weinmann06,Weinmann08,Carollo13,Carollo14}, this information is essential
in order to elucidate the relation between group environment and the
evolution of the member-galaxies.

Earlier efforts to study at X-ray representative
samples of groups based on optical selection of member
galaxies~\citep{Mahdavi97,Mahdavi00,Burns96,Ebeling94,Ponman96},
have been mostly based, however, on shallow exposure RASS data and/or on
limited information about the physical properties member galaxies.
More recently, the Complete Local Group Sample (CLoGS; O'Sullivan et al.\ 2014)
project is attempting to produce the first statistically complete survey of
galaxy groups observed in the X-ray, optical and radio wavebands,
which should greatly contribute to address the scientific questions
raised above. Using a complementary approach, we have also started the Zurich ENvironmental Study (ZENS; \citealt{Carollo13,Cibinel13a,Cibinel13b,Carollo14}), a program to build a representative sample of galaxy groups, unbiased with respect to dynamical conditions, with multi- wavelength coverage, and with fully determined properties of the member
galaxies. The aim of our program is to study the environmental impact
on the evolution of galaxies including, amongst others, the interplay
between galaxy formation and DIM.  The ZENS project, described in more
detail in Sec.~\ref{zens:se}, differs in several respects from CLoG,
which is also briefly discussed there for comparison. 

In this paper we report on initial efforts to study the DIM with X-ray observations of
ZENS groups. In particular we present data taken during the past few
years with the XMM-$Newton$ and $Chandra$ telescopes on a sub-sample
of ZENS groups.  In this paper we focus on the diffuse X-ray emission while 
we report on the X-ray point-sorce detections in a companion paper~\citep{Silverman2014}
aimed at investigating the role of AGN in the context of galaxy evolution in groups.

The XMM-$Newton$ and $Chandra$ observations,
including the data analysis, are described in Sec.~\ref{obs:se}, while
results concerning the diffuse X-ray emission are presented in
Sec.~\ref{diffuse:se}.  Results and conclusions are summarized in
Sec.~\ref{conc:se}. In the following we assume a concordance
$\Lambda$-CDM universe with parameters $\Omega_m=0.2792$,
$\Omega_b=0.0462$, $\Omega_\Lambda=1-\Omega_m$, and Hubble constant
$H_0=70$ km s$^{-1}$ Mpc$^{-1}$~\citep{komatsuetal09}.

\section{The Zurich ENvironmental Study (ZENS)}\label{zens:se}

The ZENS project, including its design, the observations and the
publicly available catalogue, and all environmental, structure and
photometric measurements, are described in great details in the
first three papers of the ZENS series
(\citealt{Carollo13,Cibinel13a,Cibinel13b}, respectively Paper I, II
and III), which we refer to, for further information.  In the
following we summarize aspects of the project which are relevant to 
this paper.

The ZENS data set consist of 141 groups, containing a total of 1630
galaxies, randomly extracted from the complete sample of groups of the
2dF Percolation$-$Inferred Galaxy Group catalog~\citep{Eke04}
(2PIGG), with at least five cataloged members and within the narrow
redshift range, $0.05 < z < 0.0585$. 

The groups are classified in terms of total mass, dynamical state
(relaxed or unrelaxed) and location within the large scale structure
environment, i.e., their proximity to massive clusters, filaments, or
voids.  

The ZENS galaxies have been fully characterised in terms of their
stellar mass, star-formation activity, morphological/structural properties  
and central/satellite rank within the group.  Resolved information such as size and strength of bars, color gradients, color maps as well as bulge and disks colors, stellar
masses and structure properties have been also derived for our ZENS
sample.

Among extant projects with similar objectives to ours, the already
mentioned CLoGS is probably the one that comes closest in scope.
In particular, the CLoGS project is aiming for a representative catalog
of 53 local groups ($z<0.02$) with
optical, radio and XMM or Chandra coverage. Our ZENS study differs
from this projects in a number of important ways. 
The typical richness of CLoGS groups, $N\gtrsim$ 20, is larger than
the richness of the ZENS groups which have typically 10 members and
thus bracket smaller halo masses. 
In addition, the CLoGS group are
selected to have at least one luminous ($L_B>3\times10^{10}L_{\odot}$)
early-type galaxy.  While this likely ensures a group higher X-ray
luminosity, it also likely introduces a bias in the probed sample
towards more evolved/massive systems.  
Furthermore, while a basic morphological classification and
integrated photometric information is also available in the HYPERLEDA
catalog \citep{Paturel_et_al_2003} from which CLoGS is based, the ZENS
dataset improves on such data by providing, as described above a comprehensive set of detailed properties for the whole galaxies, their
bulges and disks.

\section{Observations: Data Acquisition, Analysis and Source Detection} \label{obs:se}

\subsection{XMM-$Newton$}

We have carried out XMM-$Newton$ observations of ZENS groups during
three consecutive observational programs (PI: Miniati, Prop. \#
065530, Prop. \#067448, Prop. \#069374). As of today, 9 observations
of groups have been performed.  We have also retrieved data for 4
additional groups from the archive, for a set of 13 XMM-$Newton$
observations of our ZENS groups.

Details of the observations are presented in Table~\ref{xmm_sample},
which reports: (1) the target name, (2) the observation ID, (3-4) the
RA, DEC coordinates in J2000, (5) the group redshift, (6) the number
of member galaxies, (7) the group mass inferred from the stellar mass,
(8) XMM cycle of observation, (9) and the exposure time.  The last
column (10) contains notes about the observational data, in particular
whether the observations suffered from ``flares''.

The exposure times were estimated based on the expected X-ray
luminosity of our targets, in the range $3\times 10^{40} <L_{\rm 0.5-2
  keV}$/(erg s$^{-1})<$5$\times10^{41}$~\citep{RasmussenPonman09}.
Detection of X-ray emission from groups is notoriously challenging.
This is partly because X-ray luminosity scales with total mass of the
emitting virialized system and becomes faint at the group mass
scale. In addition, at such mass scale the X-ray luminosity-mass
relation becomes characterized by considerable
scatter~\citep{EkcmillerHudsonReiprich12}, causing uncertainties in
the observational forecast.  While this reminds us of the risk of
observational biases in X-ray selected group samples, it also
illustrates the difficulty associated with amending the problem.

Different exposure times in different observing cycles reflect
different strategies underlying our successive proposals. In
particular, in AO-9 we used long exposures as we aimed at measuring
the thermodynamic properties of the intragroup medium.  This, however,
was achieved only for one of the three observed groups, which led us
to revise our strategy in AO-10, and split the program in two separate
steps: (1) first carry out a short exposure, to determine the group
luminosity and then follow up with deeper observations bright enough
groups that their thermodynamic properties can be determined with
reasonable exposure times. This approach is made difficult, however,
by the large incidence of flares, which can completely corrupt the
data for short exposure times. Slightly longer exposures were therefore
employed in AO-11.

The standard processing of the XMM observations included the screening
for the flares, and making a composite images from all 3 detectors in
the 0.5-2 and 2-7.5 keV bands, excluding the energies affected by the
strong instrumental lines, as discussed in~\cite{Finoguenov07}.
We only used 0.5-2 keV images for detection extended emission (hard
band is used to search for AGNs).  We used the procedure
of~\citet{Finoguenov07} with updates described in~\citet{Bielby10}.
We monitor the S/N ratio maps to control the quality of background
removal. On several occasions hot MOS1 and MOS2 chips were
removed. The flux of detected point sources was removed using PSF
model of XMM before proceeding with extended source detection, as
described in~\citet{Finoguenov09}. In case of both Chandra and XMM
observations we performed parrallel analysis and compared the results,
showing that estimated fluxes do not depend on the choice of the
instrument.

Table~\ref{xmm_extended_sources} summarizes measurements of the X-ray
emission from the diffuse intragroup medium, including upper limits
for the non-detections. In particular it reports: (1) target name, (2)
total X-ray counts (extended and point sources including background),
(3) background counts only, (4) counts due to point sources, (5) count
extraction radius, (6) estimate of the total X-ray luminosity in the
rest frame 0.1--2.4 keV band and, (7-9) corresponding virial mass,
virial radius and temperature, using the scaling relation of
\citep{Leauthaud10}, (10) the source flux extrapolated to the virial
radius, using the procedure outlined in \citep{Finoguenov07}, (11)
signal to noise ratio.  Note that the formal statistical significance
of the detected emission is larger then the reported flux
significance. The results based on these measurements are discussed in
Section~\ref{diffuse:se}.

\subsection{Chandra}

We also carried out $Chandra$/ACIS-I observations of 12 ZENS galaxy
groups in Cycle 11 (Prog. \# 11700688; 120 ksec).  The observations
were executed between September 2009 and October 2010.  Each target is
observed for 10 ksec, the main objective of these observations being
the detection with at least 4 counts in the broad
energy band 0.5-8 keV, of AGNs at $z\sim0.05$ down to a
limiting luminosity of $L_{(0.5-8~{\rm keV})} \sim 4\times 10^{40}$
erg s$^{-1} $.  The field-of-view of ACIS-I ($16.9\arcmin
\times 16.9\arcmin$; CCDs \#0-3) is sufficient to cover the sky area
of these galaxy groups (size $<10\arcmin$).  
The target positions were chosen to avoid having galaxies falling
within or near chip gaps; this was accomplished by adjusting the
pointing location once the planned observation date was set thus the
roll angle was known.  In Table~\ref{chandra_sample}, we provide
details on the individual exposures.

We employ a method to detect extended emission in our $Chandra$
observations similar to that of a number of recent analysis
\citep[e.g.,][]{Cappelluti12,Tanaka12b}.  We screen the
event file by removing time intervals affected by flares, using the
CIAO $lc\_clean$ tool.  Exposure maps (effective area versus sky
position) are generated for each observation listed in
Table~\ref{chandra_sample} using $mkinstmap$ and $mkexpmap$.  To do
so, we generate an instrument map (effective area versus detector
position) in the 0.5-2 keV band using a model weighing scheme of a
power law distribution with index of 1.7.  For our purpose, the
difference in the final exposure map when using either a power-law or
thermal spectrum is negligible.  Using the merged event files, we
generate a combined image for each galaxy group in the 0.1-2.4 keV
band.

The detection of diffuse emission requires the removal of point
sources.  To do so, we run $wavdetect$ on the full image in the
0.1-2.4 keV band with a significance threshold of $5\times 10^{-6}$.  An input
parameter of relevance is the encircled energy (39.3\%) of the PSF at
an energy of 1.5 keV.  In addition, elliptical source regions of the
output detections have a size of 3$\sigma$ of the PSF.  After excising
the photon events in these regions, an artificial background level is
applied based on the level in a larger background region local to each
source detection.

The background X-ray emission is estimated using the procedure of
\citep{Hickox06} that uses the particle background (of
non-astrophysical origin) measured from the stowed position of ACIS
(http://cxc.cfa.harvard.edu/contrib/maxim/acisbg) in the 9.5-12 keV
energy range.  We then scale this background map to match that
expected in our observations by the ratio
$C_{data,9.5-12}/C_{stow,9.5-12}$ where $C_{data}$ are the counts
measured in the data while $C_{stow}$ are the counts detected in the
stowed position.

We run the 'wvdecomp' algorithm \citep{Vikhlinin98} on the data after subtracting the
smoothed background maps, to search for extended sources on angular
scales exceeding 30 arcseconds.  In four cases, we detect emission and
use a circular region with a radius matched to the extent of the
emission for flux estimates. In cases where no emission is detected,
we use a circular region with a 2-3 arcminute radius to determine an
upper limit (2$\sigma$) on the flux.  We also measure a `control' flux
estimate in the source-free zone to refine the background subtraction.
In cases of positive signal, we subtract it from the source emission
with a scaling based on flux extraction area.  We provide the
following specific measurements of diffuse emission: (1) total counts,
(2) background counts, (3) counts associated with embedded point
sources, (4) the extrapolated flux of the source, using the procedure
outlined in \citep{Finoguenov07}, (5) estimate of the total X-ray
luminosity in the rest frame 0.1--2.4 keV band and (6) a corresponding
virial mass using the scaling relation of \citep{Leauthaud10}.  All
results based on these measurements are discussed in
Section~\ref{diffuse:se}.

In Table~\ref{chandra_extended_sources}, we provide our measurements
of the X-ray emission from the diffuse intragroup medium, that include
upper limits for the non-detections.  As for table relative to the
XMM-$Newton$ diffuse emission, the columns are as follows: (1) Name,
(2) total X-ray counts (extended and point sources including
background), (3) background counts only, (4) counts due to point
sources, (5) count extraction radius, (6), luminosity, (7-9)
$M_{200}$, $R_{200}$, and X-ray temperature, respectively, based on
the scaling relation of \citep{Leauthaud10}, (10) source flux, and
(11) significance of the detection of extended emission.

\subsection{Selection Bias} \label{bias:se}

Since the groups were selected to be within a specific mass range, the
mean properties of the groups computed a posteriori as a function of
the selection parameter are affected by a bias which need to be taken
into account. The bias is due to the fact that systems below the mass
threshold will still be selected due to statistical errors in the mass
estimates, while systems above the mass threshold will not be selected
due to mass underestimates. The effect is not symmetric across the low
and high mass thresholds, because the mass function is a steep
power-law.  To amend for this effect we
follow~\citet{Vikhlinin09} who suggested a way to estimate the bias
factor for the purpose of optimizing the measurements of cluster
scaling relations.  In their approach the bias correction is specific
to the individual values of the selection parameter $M$, and further
depends the threshold, $M_{th}$, and dispersion, $\sigma_M$, of the
latter as
\begin{equation}
\beta(M,\sigma_M)=\frac{\int_{M_{th}} (M-M_{th})
  e^{-\frac{(M-M_{th})^2}{2\sigma_M^2}} dM }{ 
  \int_{M_{th}} 
  e^{-\frac{(M-M_{th})^2}{2\sigma_M^2}} dM }.
\end{equation}
The value of $\sigma_M$ can be determined from the data and to first
order is unbiased~\citet{Vikhlinin09}.  In
Fig.~\ref{lx:fig} presented below we report bias
corrected estimates of the group masses, assuming that $\sigma_M$
corresponds to 50\% of the mass measurements dispersions.

\section{Results} \label{diffuse:se}

Our X-ray observations of galaxy groups in ZENS with XMM-$Newton$ and
$Chandra$ enable us to begin the construction of a sample selected
without an obvious preference on their X-ray properties or dynamical
state.

Of the 13 ZENS groups observed with XMM-$Newton$ and listed in
Table~\ref{xmm_sample}, one third have data that is  severely
affected by flares,  and is thus not usable. This problem affected
equally all observations with exposure time below 40 ksec. Of the
remaining 9 groups with valid data, X-ray emission from the DIM was
clearly detected in 5 cases, with a typical signal-to-noise
4--5, in addition to a two sigma detection. 
Group 2PIGG\_s1571 was an exception, with a signal-to-noise
46.87, much higher than for the other groups.  The reminder of the
X-ray observations includes a 1.8-$\sigma$ detection, a 2-$\sigma$
upper limits and an upper limit.

In comparison, of the 12 ZENS groups observed with $Chandra$ and
listed in Table~\ref{chandra_sample}, X-ray emission from the DIM was
detected in a total of three cases. One of these cases corresponds to
2PIGG\_s1571 and, as for the case of XMM-$Newton$, it is detected with
a high signal to noise. The flux measured by the two telescopes is
consistent within the reported statistical error. The reminder of the
X-ray observations include a 2-$\sigma$ detections, while the rest are
upper limits.

These results reveal the presence of diffuse intragroup gas inside the
shallow potential well of small scale groups with masses around
10$^{13}M_\sun$. The X-ray emission is rather faint, with typical flux
$\le 10^{-14}$ erg cm$^{-1}$ s$^{-1}$, i.e. one order of magnitude
fainter than typical ROSAT groups (RASS).  In addition, for several
groups we were able to measure  upper limits, which implies that
we are mostly probing the upper envelope of the X-ray emitting,
optically selected ZENS groups.  While this is likely due to diversity
in X-ray luminosity of our groups at fixed mass, we cannot completely exclude that at least in some cases the effect is 
enhanced by uncertainties in the mass determination of the
groups.

\subsection{X-ray vs Optical Images}

\begin{figure}
\centering
\includegraphics[width=0.5\textwidth]{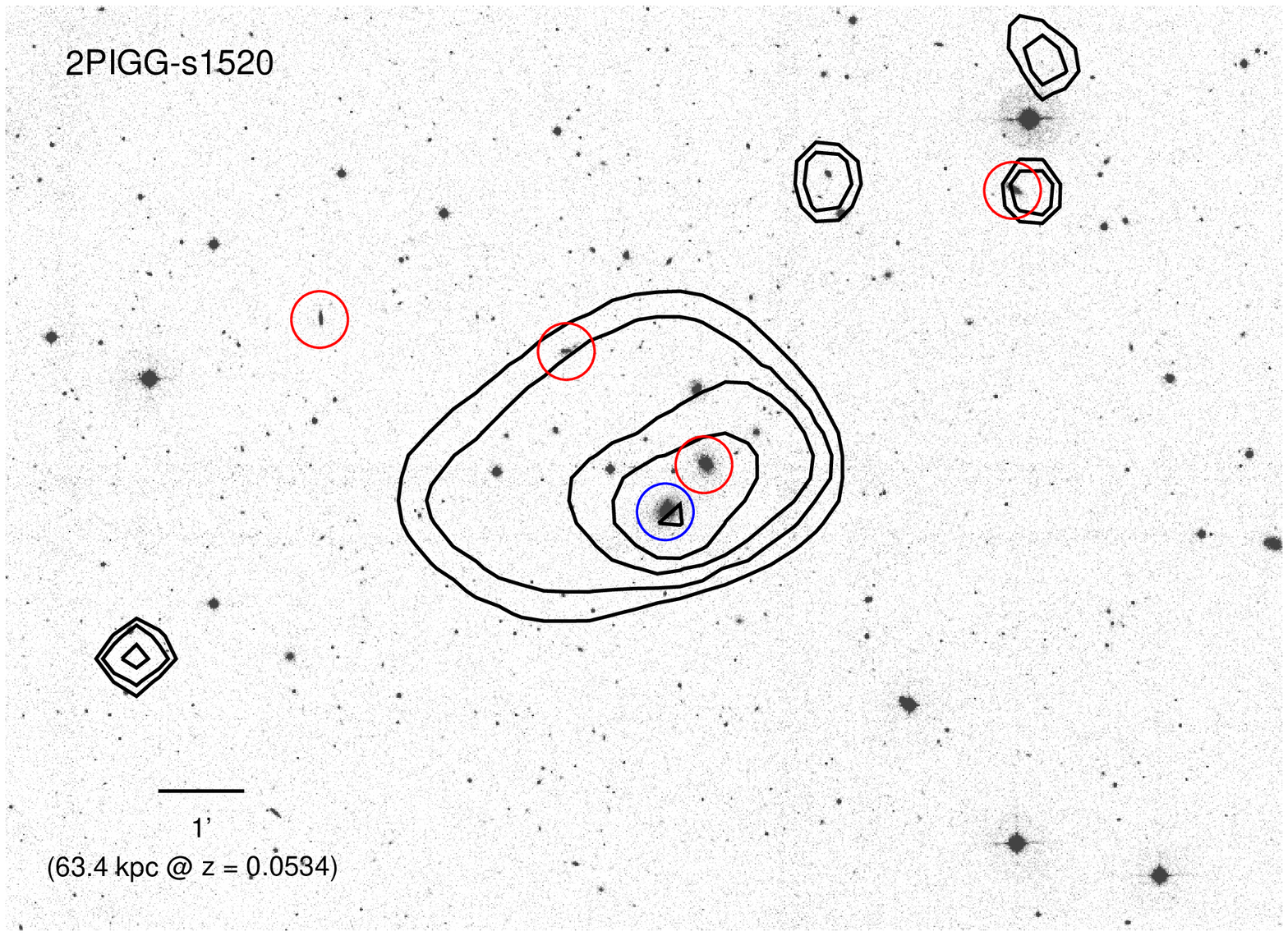}\includegraphics[width=0.5\textwidth]{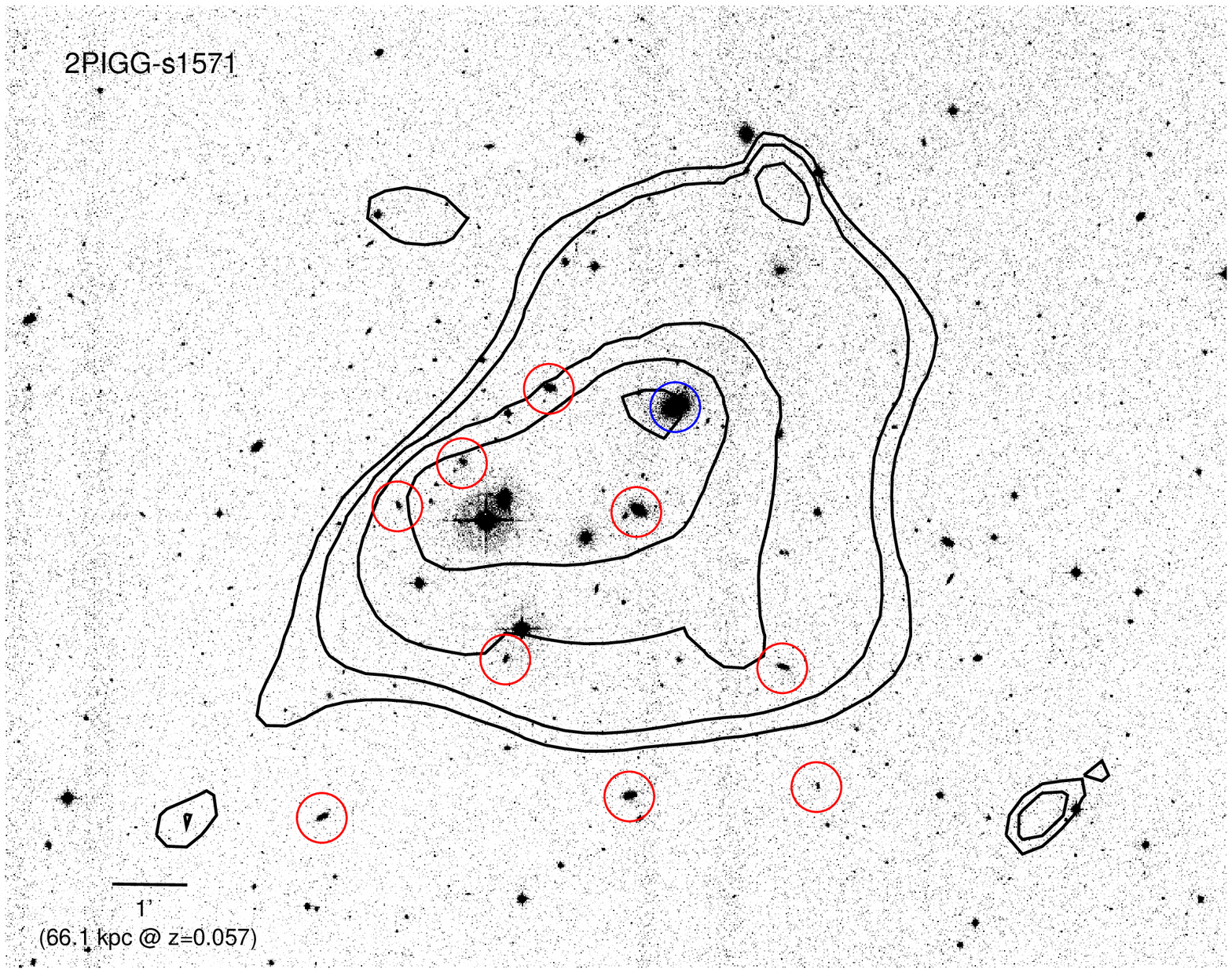}
\includegraphics[width=0.5\textwidth]{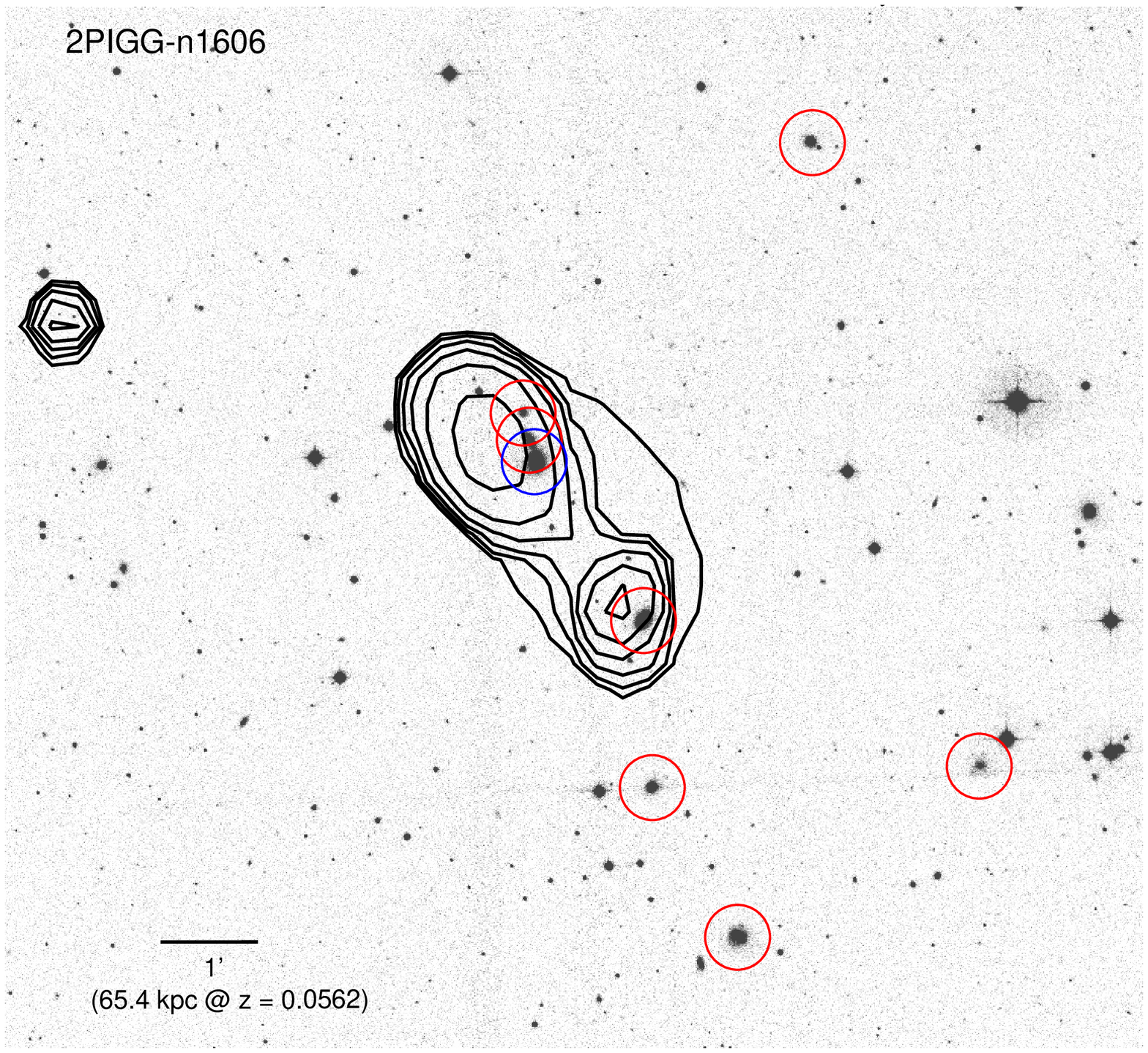}
\caption{XMM-$Newton$ measurement of the Diffuse Intragroup Medium
  X-ray emission from the ZENS groups 2PIGG\_s1520 (top) 2PIGG\_s1571
  (centre) and 2PIGG\_n1606 (bottom), overlaid with optical i band
  image.  Black contours indicate the surface brightness level of
  X-ray emission on a log scale.  Red circles mark the galaxy members
  and the those in blue the central galaxy determined prior to
  knowledge of the X-ray emission.}
\label{xmm:fig}
\end{figure}

\begin{figure}
\centering
\includegraphics[angle=0,scale=0.4]{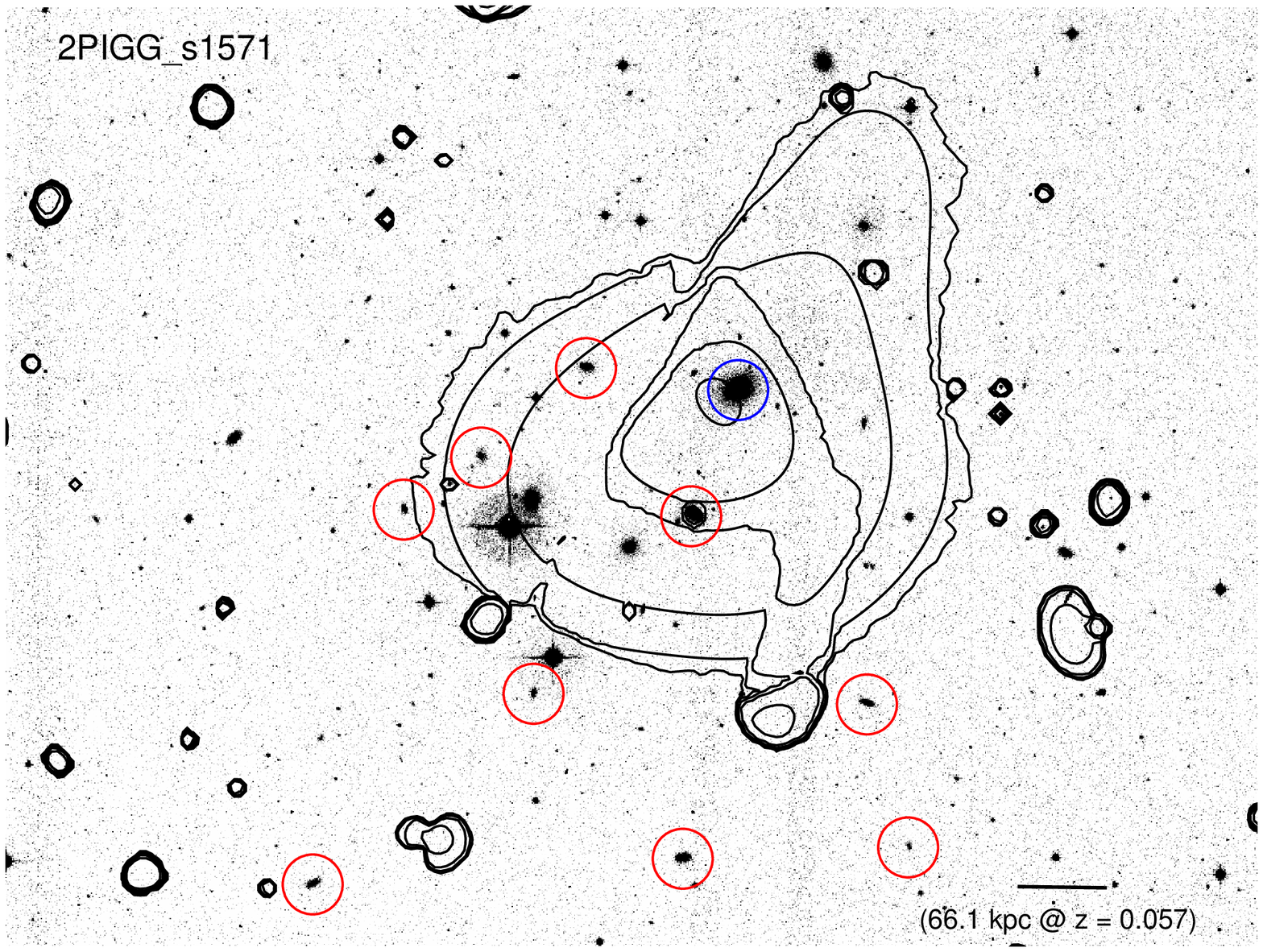}
\includegraphics[angle=0,scale=0.4]{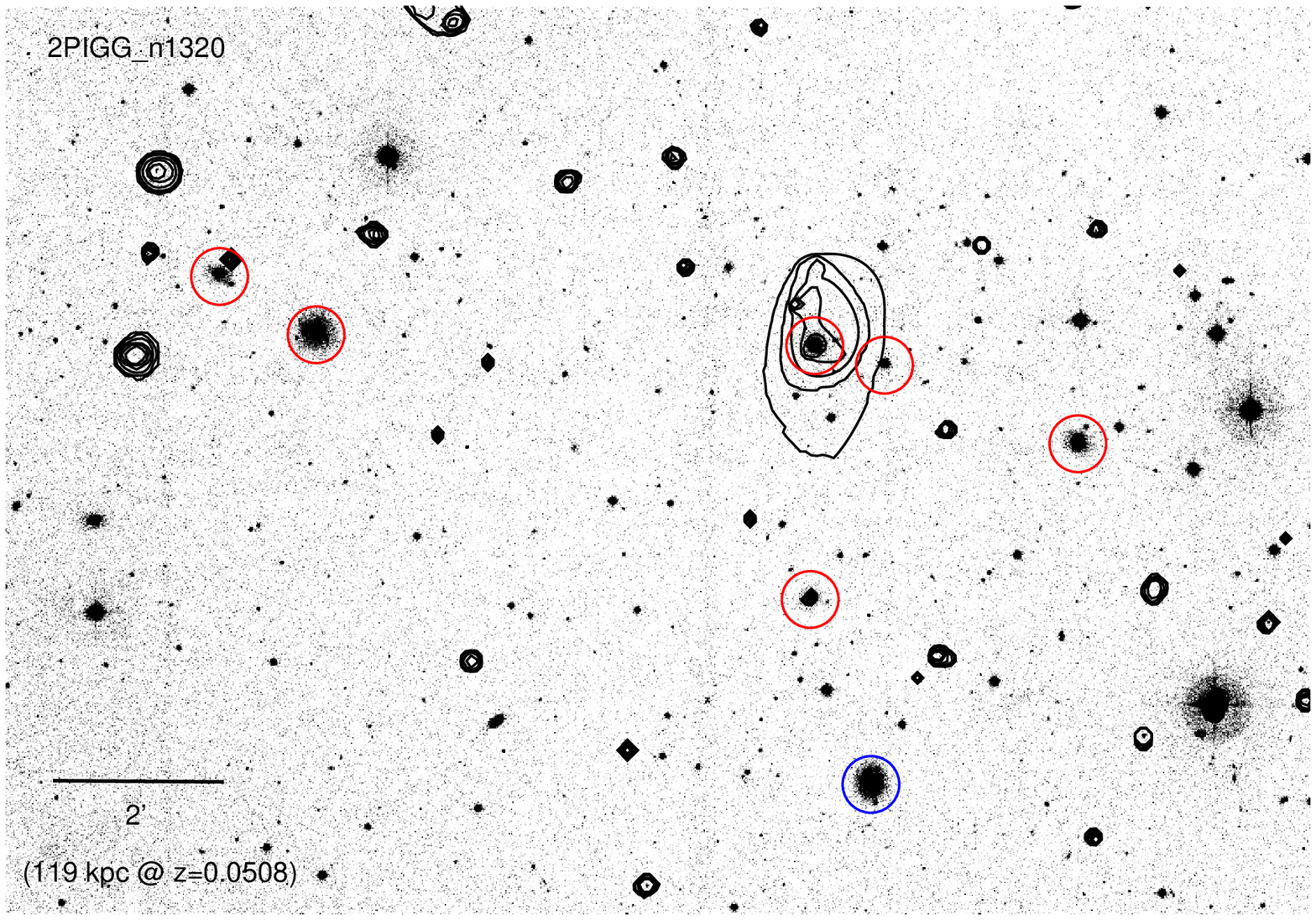}
\caption{Chandra measurements of the Diffuse Intragroup Medium 
X-ray emission from the ZENS groups 2PIGG\_s1571 (top) and
  2PIGG\_n1320 (bottom), overlaid with optical i band image.  Black
  contours indicate the surface brightness level of X-ray emission on
  a log scale.  Red circles mark the galaxy members and the those in
  blue the central galaxy determined prior to knowledge of the X-ray
  emission.}
\label{chandra:fig}
\end{figure}

In Fig.~\ref{xmm:fig}--\ref{chandra:fig} we show examples of detection
($S/N>3$) of diffuse X-ray emission from ZENS groups, based on both
XMM-$Newton$ (2PIGG\_s1520, 2PIGG\_s1571, 2PIGG\_n1606) and $Chandra$
(2PIGG\_s1571, 2PIGG\_n1320) observations.  The contours indicate the
surface brightness level of the X-ray emission on a log-scale and are
superposed to optical i-band images of the respective group fields.
Member galaxies are identified with a small circle, red in the case of
satellites and blue for central galaxies. The latter correspond to
galaxies with the highest stellar mass that also satisfy the following
requirements: (1) to be close in position to the group centre and, (2)
to move with respect to the group bulk at a velocity that is within
one $\sigma$ of the group velocity dispersion~\citep[see][for further
details]{Carollo13}.  In our ZENS work, we  consistently assumed that the so-defined central galaxies identify the centres of the groups. This approach differs from the original 2PIGG procedure of Eke et al.\, who computed and used, as  group centre,  a weighted average of the galaxy positions. Both these approaches were developed
prior to any knowledge of the groups' X-ray emission, which now allows for a
sensible consistency check of either optical definitions of centres, and between group properties determined  with optical and with X-ray data. Of particular importance is the fact that the precise definition of a group centre also establishes the classification of its dynamical state.

 We  thus inspect  the composite images in
Fig.~\ref{xmm:fig}--\ref{chandra:fig}.  Furthermore, Table~\ref{matches:tab} lists the difference between the
2PIGG optically-averaged and the X-ray centres of the groups (3rd column),
as well as the distance of the group central galaxy to the 2PIGG 
(4th column) and X-ray (5th column) centres.  The first three columns
contain to the groups name and optical equatorial coordinates, while
the last column indicates the groups' dynamical state, according to
the optical data.  The Table includes all groups for which a group
centre could be reliably determined from the X-ray data.  

For three out of the six listed groups (2PIGG\_s1520, 2PIGG\_s1571,
2PIGG\_n1320), the 2PIGG and X-ray centres coincide to within a few
arc seconds.  We furthermore note that for two of these three groups, namely 2PIGG\_s1520 and 2PIGG\_s1571, the 2PIGG centres  coincide with the central galaxy. 
The group 2PIGG\_s1571 was classified in Paper I as a relaxed group, with the central galaxy identification indeed  in agreement with both the 2PIGG-averaged centre and the 
peak of the X-ray emission. This group thus provides reassurance that, in well-behaved systems, the centre and central galaxy associations based on optical data are robust and physically-motivated.    

Things can be trickier however for more complex structures.  The group 2PIGG\_s1520 is  classified as unrelaxed based on the optical data, but nevertheless there is a good agreement between the X-ray peak and the 2PIGG-averaged centre.  
On the other extreme,  the group 2PIGG\_n1606 was classified   as relaxed according to the optical data alone, but  its diffuse emission is
clearly double peaked, which is suggestive of a merger event. 
 In 2PIGG\_n1606 the offset between the optical and X-ray
centres as well as optical centre and central galaxy, is much larger
than the separation between X-ray centre and central galaxy. This is
consistent with the X-ray picture that the system is undergoing a
merger, the system is unrelaxed, and the X-ray emission traces the
largest of the merging structures. 
In the case of 2PIGG\_1320, which is also
an optically relaxed group, the galaxy corresponding to the peak of
the X-ray emission is not the same as the optically-identified  central galaxy of the group.

The remaining   groups reported in Table~\ref{matches:tab} also show
some degree of mismatch between the optical and X-ray centres. The group 2PIGG\_n1746, characterized by the smallest discrepancy, is  classified as
relaxed from the optical.   2PIGG\_n1746 is 
characterized by a mild separation between the 2PIGG and the X-ray
centres, although the X-ray centre is closer to the
central galaxy than to the optical centre. Therefore this group is
also likely to be also dynamically unrelaxed in spite of the optical
classification, as in the case of 2PIGG\_n1606 discussed above.
In the case of 2PIGG\_n1377, classified as dynamically unrelaxed, the separation between the optical and X-ray centre is mild, while the separation of either
centre to the central galaxy is quite large.

These examples illustrate that there is a substantial difficulty in identifying group centres from optical data alone, and, most likely, also a substantial diversity in
group properties and dynamical conditions. While the optical and X-ray
data are often consistent, in several cases important discrepancies
become apparent and complementary X-ray observations become important
for a complete picture~\citep[see, also][]{George12}. It is 
the combination of high quality optical data and X-ray observations
that reveals subtle properties of the dynamical state of groups, which would
be missed by an analysis based purely on either data set alone.
This information is important in studies of galaxy properties and
evolution as a function of group environment, which aim at
establishing the physical processes and conditions responsible for
triggering evolutionary mechanisms.  Examples include structural and
morphological properties as well as star formation and or AGN activity
as a function of either the group-centric distance or the group
dynamical conditions. It is obvious that mis-classifications of the
group centre and/or dynamical state, introduce noise in the observed
relations which can be amended with  X-ray data.
This straightforward comparison that we have conducted is therefore, above all,   a strong warning of the relevance to have complementary  X-ray information in order  to establish the full dynamical picture of a galaxy group.

\subsection{L$_X$ vs M Diagram and Mass Determinations} \label{mass:se}

\begin{figure}
\centering
\includegraphics[angle=0,scale=0.4]{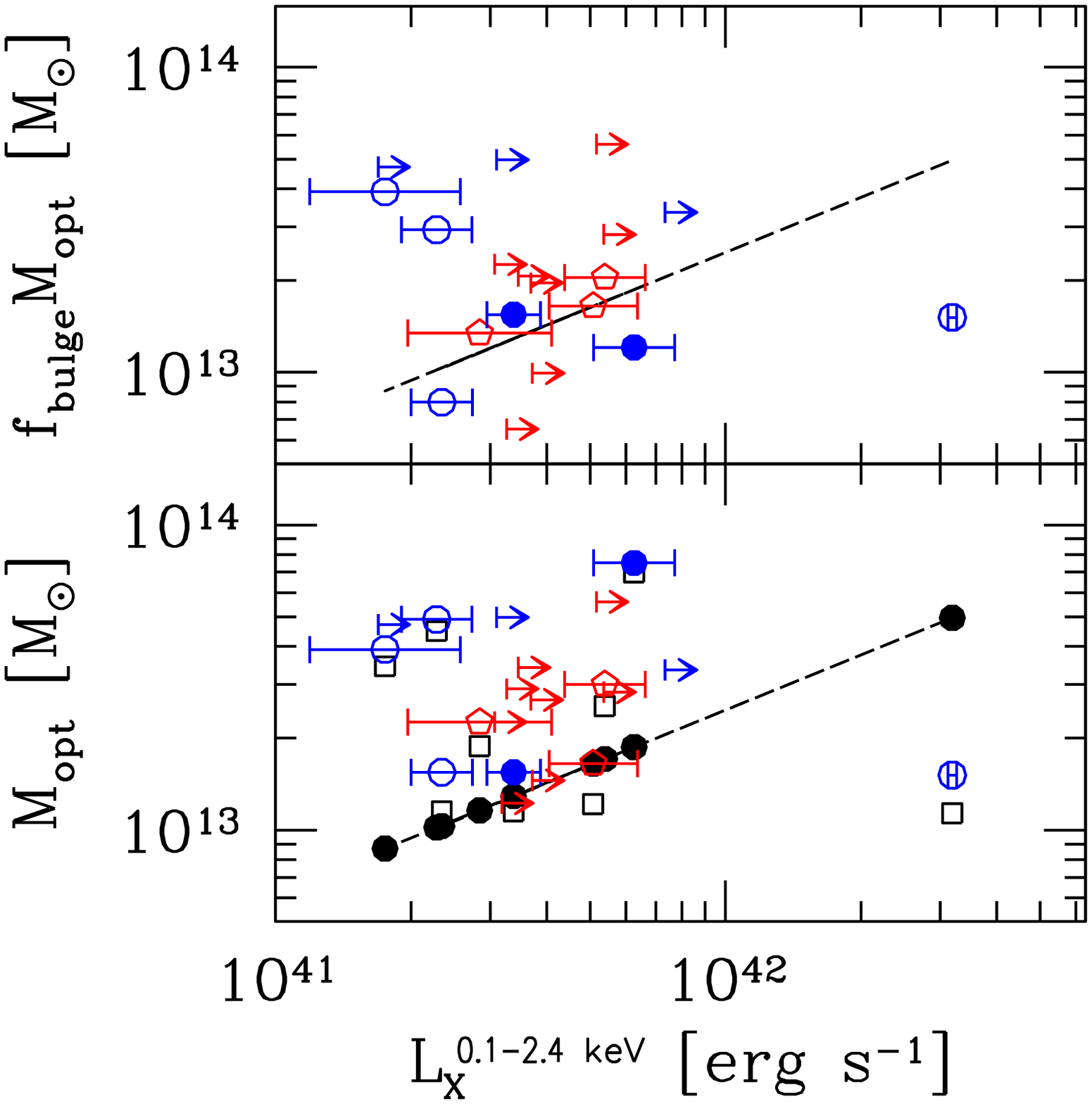}
\includegraphics[angle=0,scale=0.4]{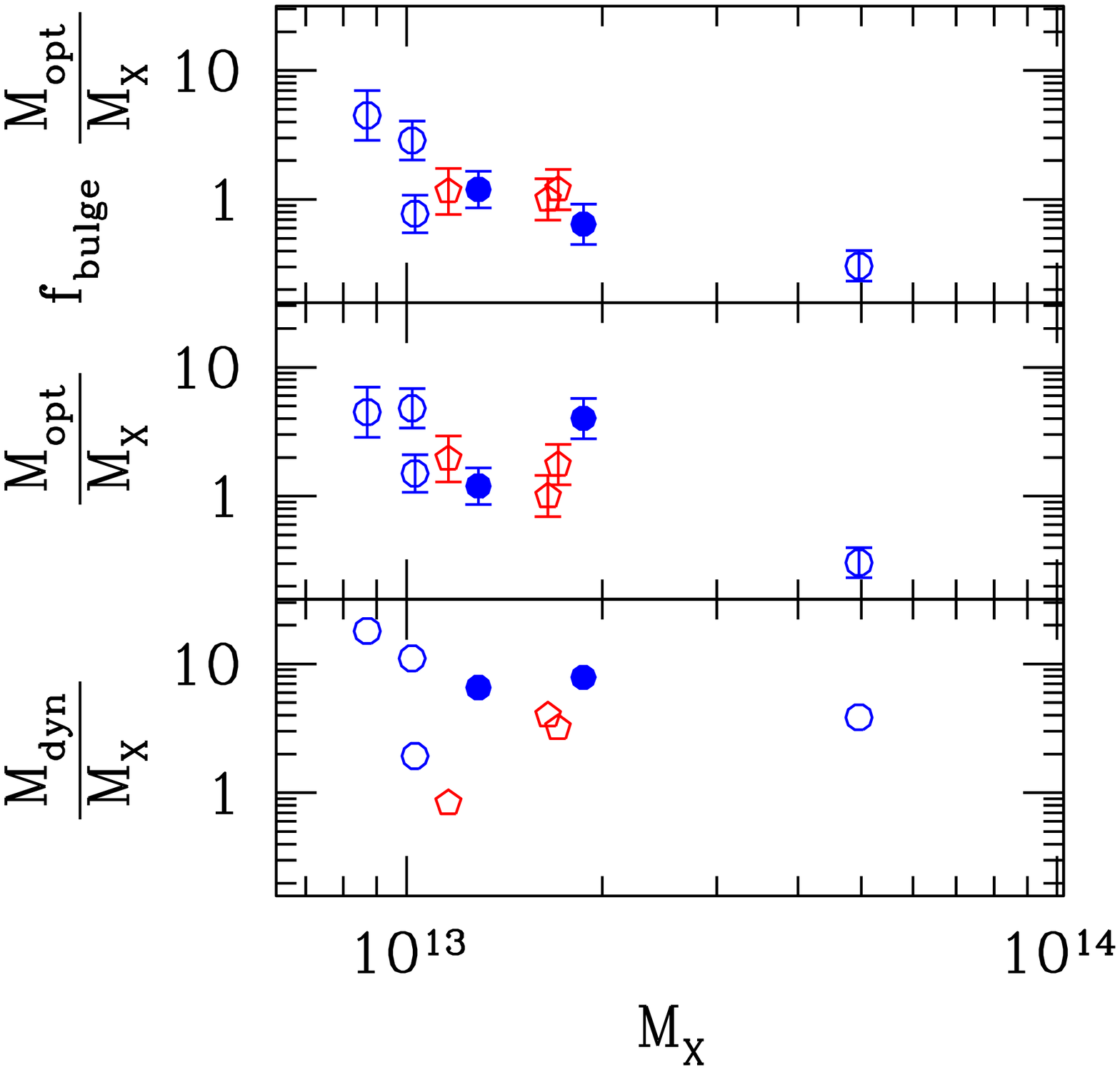}
\caption{{\it Left:} Group mass versus X-ray luminosity from the diffuse intragroup
  medium. Measurements from XMM-$Newton$
  (blue circles) and $Chandra$ (red pentagons) with signal-to-noise
  better than two are shown together with upper limits for the
  respective instruments. Open and solid symbols denote relaxed and
  unrelaxed system, respectively.  Horizontal bars correspond to
  statistical errors in the X-ray luminosity as reported in
  Table~\ref{xmm_sample} and~\ref{chandra_sample}. The group mass for
  the ZENS sample is obtained from the groups optical luminosity and
  is given in the same Tables.  For comparison, the black solid
  circles connected by a dash line indicate the position of the groups
  in the diagram if their mass is derived from the X-ray luminosity
  using the $L_X-M_X$ relation established from the joint analysis of
  X-ray and weak lensing for groups in COSMOS~\citep{Leauthaud10}. The
  black open squares are the optically determined groups masses, $M_{opt}$, 
  corrected for a selection bias as discussed in Sec.~\ref{bias:se}.  {\it Right:}
  dynamical (bottom), optical (middle) and bulge-based optical
  ($\equiv f_{\rm bulge}\times M_{\rm opt}$) group mass estimates
  divided by the X-ray mass. Data from XMM-$Newton$ (blue circles) and
  $Chandra$ (red pentagons) with signal-to-noise better than two are
  shown. Open and solid indicate relaxed and unrelaxed systems
  respectively. Error bars include errors associated with X-ray mass
  determination reported in Table~\ref{xmm_extended_sources}
  and~\ref{chandra_extended_sources}, and 30\% mass estimate for the
  optical mass~\cite[see][]{Carollo13}.  }
\label{lx:fig}
\end{figure}

In the bottom left panel of Figure~\ref{lx:fig} we plot the group mass
as a function of the X-ray luminosity from the diffuse intragroup
medium. Data points with signal-to-noise better than two are shown for
XMM-$Newton$ and $Chandra$ as blue circles and red pentagons,
respectively. Data points with lower significance are also shown as
upper limits for each respective instrument. Open and solid style
indicates relaxed and unrelaxed systems, respectively, while horizontal
bars correspond to statistical errors in the X-ray luminosity reported
in Table~\ref{xmm_sample} and~\ref{chandra_sample}.  The group mass,
$M_{opt}$, listed also in the above Tables, is provided by the ZENS
catalog and is based on the optical luminosity of the group member
galaxies. It is computed from the total optical luminosity
of the group $L_{Group}$, using the mass-to-light ratio~\citep[see][for
details]{Eke04,Carollo13}
 
\begin{equation}
  \log\left(Y_{b_j} \right) = 2.28 + 0.4 \tanh\left\{1.9\left[\log\left(L_{Group}\right) - 10.6\right]\right\}.
\end{equation}
 
The black open squares correspond to the group masses, $M_{opt}$,
corrected for a selection bias as discussed in Sec.~\ref{bias:se}.

In order to compare with prediction from scaling relations, we also
present in the same plot the black solid circles connected by a dash
line. They indicate the position of the groups in the diagram if we
derive their mass, $M_X$, from the X-ray luminosity using the
$L_X-M_X$ relation in~\citet{Leauthaud10}.  This relation is based on
a joint analysis of X-ray and weak lensing of groups in COSMOS survey.
The plot shows that our groups depart significantly from the scaling
relation inferred from the X-ray selected groups. Except for
2PIGG\_s1571, our ZENS groups appear to be under-luminous for their
stellar content.  This is not surprising and accords to finding from
previous studies~\cite[most recently][but see also discussion in the
Introduction]{Connelly12}, which have shown that optically selected
groups and clusters of galaxies tend to be less X-ray bright than
X-ray selected ones. Corrections for the selection bias (open black
squares) cannot account for the discrepancy.

The discrepancy is further illustrated in the middle-right panel,
where the optical to X-ray mass ratio is plotted as a function of the
X-ray mass, $M_X$.  There is a clearly relative bias between the two
estimates, with the optical masses being always larger than the X-ray
masses (with a single exception represented by 2PIGG\_s1571). The
discrepancy becomes significantly larger when the X-ray mass is
compared to the dynamical mass (bottom right panel), $M_{dyn}$,
determined from dynamical arguments~\cite[i.e.][]{Connelly12}
 
\begin{equation}
M_{dyn}=\frac{3\sigma_v^2R_{200}}{G_N},
\end{equation}
 
indicating that the latter is a considerably less reliable mass proxy,
due to large uncertainties mostly associated with the determination of  
velocity dispersions with respect to the group
centre of mass from a few  member galaxies. 

The optical mass determinations ($M_{opt}$) are usually thought to overestimate the
actual mass of the group (see Sec.~\ref{mass:se} below).
In fact, the above bias is typically ascribed to the dynamical conditions
of the group. However, in our limited sample 
there seems to be no indication that unrelaxed groups
(solid symbols) are more under luminous than relaxed ones (open
symbols).  In addition, inspection of the individual group properties
show that the under luminous groups span the full range in terms of
bulge fraction of the central galaxy, which suggests that the under
luminosity may not be simply related to a recent formation of the
group.  We have found, however, a slight correlation between the ratio
of the optical to the X-ray mass, and the bulge fraction, in the sense that the discrepancy between the optical
and X-ray masses seems to increase towards smaller bulge
fraction. This suggests that the group stellar mass in the bulge component
is a better proxy to the group mass than $M_{opt}$ based on the total optical emission.
This is supported by the results shown in the top-left panel of
Fig.~\ref{lx:fig}, where we multiply the ratio of the optical to X-ray
group mass times the bulge fraction and plot it as a function of the
X-ray mass.  Compared to the plot of the original mass ratio in the
middle-left panel of Fig.~\ref{lx:fig}, the discrepancy
between the two mass determinations appear significantly reduced,
although not completely removed.  Note that a similar results were also found
in~\cite{Andreon12}, although at considerably larger mass scales. 
We have further found that, for the small sample of
groups considered here, the above correction appears to be unbiased with
respect to the different values of the bulge fraction.
Even after the correction, however, the 
distribution of the X-ray luminosities with respect to the
dash line remains asymmetric~\citep[see also][]{Connelly12},
which may suggest that the
under-luminous character of the observed ZENS groups may be genuine
to some extent.
However, the number of X-ray detections is currently  too limited to draw any
firm conclusion from the current data.

Note that, group 2PIGG\_s1571, which is detected with the best
significance, departs from the scaling predictions in the opposite
direction, i.e. is over-luminous by an order of magnitude. However,
this could well be consistent with the observed large scatter in $L_X$
at these mass scales~\citep{EkcmillerHudsonReiprich12}.

\subsection{Stellar Mass Fraction} \label{fb:se}

\begin{figure}
\centering
\includegraphics[angle=0,scale=0.75]{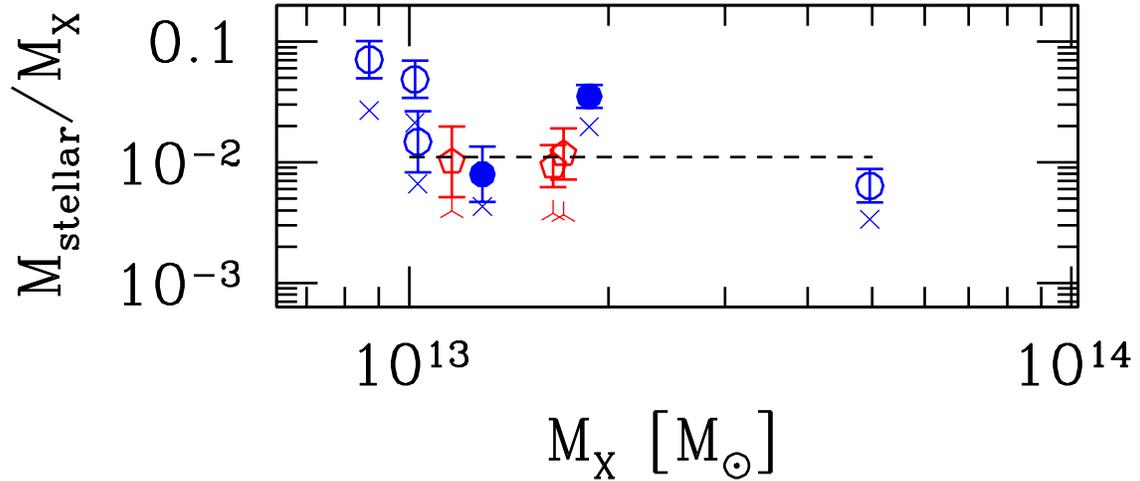}
\caption{Stellar-to-total mass ratio as a function of the X-ray mass,
  where the total mass is the X-ray mass obtained using the scaling
  relations in ~\cite{Leauthaud10}. Blue and red symbols
  indicate X-ray data from XMM-$Newton$ and $Chandra$, respectively,
  while open and solid indicate relaxed and unrelaxed systems
  respectively.  The dash line indicate the median value of 0.011.
  Only groups with X-ray detections with signal-to-noise better than
  two are shown.  Upper limits do not contribute additional
  information as they mostly crowd above the median bar of the top
  panel. Error bars include errors associated with X-ray
  mass determination reported in Table~\ref{xmm_extended_sources}
  and~\ref{chandra_extended_sources}, and errors based on the upper
  limits of the best stellar mass estimate given
  in~\citet{Carollo13}.  }
\label{f4:fig}
\end{figure}

In Fig.~\ref{f4:fig} we plot the stellar-to-total mass ratio as a
function of the X-ray mass, $M_X$, where the total mass is $M_X$.
The stellar mass is conventionally computed using galaxies with masses down to $10^{10}
M_\odot$~\citep{Leuathaud12}, where the ZENS catalogue is complete for any spectral type.
Therefore, our values of the stellar mass fraction should be treated as lower limits.
In Fig.~\ref{f4:fig} blue and red symbols indicate X-ray
data from XMM-$Newton$ and $Chandra$, respectively, while open and
solid indicate relaxed and unrelaxed systems respectively.  Only
groups with X-ray detections with signal-to-noise better than two are
shown, and upper limits are not plotted as they do not contribute
additional information since they mostly crowd above the median
values. The latter is indicated by a dashed line, which correspond to
a value of  0.011.

The stellar fractions computed in the top panel are within 30\% of the
median value, except for 2PIGG\_1571.  The median of 0.011 is
consistent with results obtained by~\citet{Connelly12} for optically
selected groups and also~\cite{Balogh11} at similar sample size.  
Our fractions are lower than
the values reported in~\cite{Giodini09}, who however measured the
stellar mass fraction within $R_{500}$, as opposed to the $R_{200}$ as
in our case, as well as~\cite{Gonzales07} who in addition took into
account the contribution from intracluster light.

For comparison we also plot the stellar mass fraction contributed by
the group most massive galaxy. It is indicated with a blue or red
cross below the circle or pentagon symbol of the corresponding group.
The plot shows that the most massive galaxy typically contains between 30
to 50 \% of the total stellar mass fraction in the group.

\subsection{$L_X$ vs Global Galaxy Properties}

As a first attempt to  investigate whether either  the diffuse X-ray emission or  the
galaxy properties or both show hints for a substantial
interaction between the DIM and the group galaxies, we searched for correlations
between the X-ray properties of the groups, i.e. their luminosity, and
the general properties of the group member galaxies as a whole.  

\begin{figure}
\centering
\includegraphics[angle=0,scale=0.75]{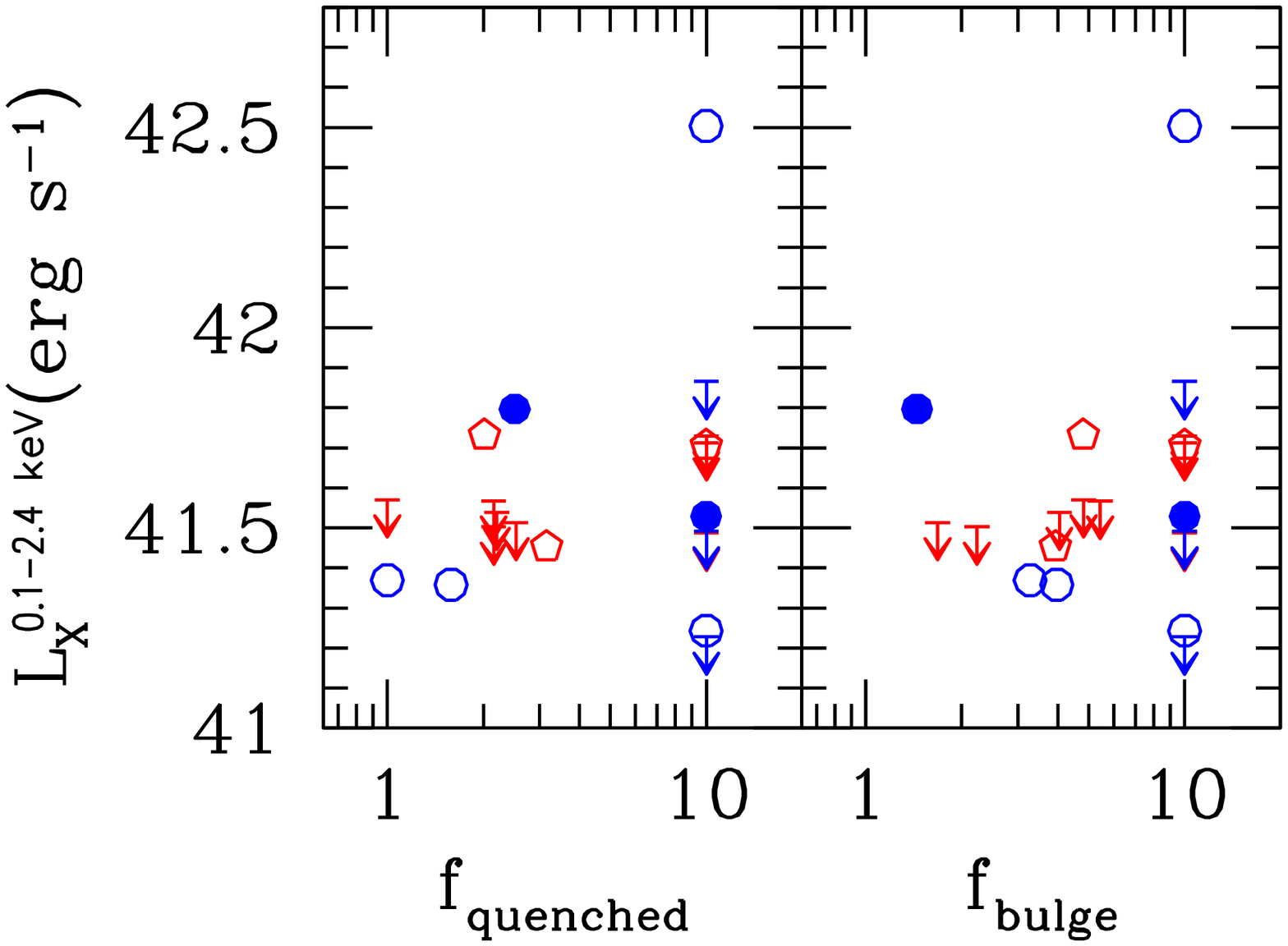}
\caption{Group X-ray luminosity as a function of the fraction of group
  galaxies with quiescent star formation (left) and the fraction of
  bulge stellar mass of the group member galaxies (right). Quantities
  are computed from galaxies with mass above $10^{10}$M$_\odot$, where
  the ZENS catalogue is complete.  The star formation for individual
  galaxies is determined through photometric and spectroscopic
  measurements and proper correction for
  incompleteness; the bulge component is determined
  using a bulge-disk decomposition based on both I and B band
  photometric data~\citep{Cibinel13b}.  Data from XMM-$Newton$ (blue
  circles) and $Chandra$ (red pentagons) with signal-to-noise better
  than two are shown together with upper limits for the respective
  instruments.  Open and solid indicate relaxed and unrelaxed systems
  respectively.  }
\label{f5:fig}
\end{figure}

In Fig.~\ref{f5:fig}
we show two examples of correlation that could in
principle reveal such an effect.  In the left panel of
Fig.~\ref{f5:fig} for each
group we plot the X-ray luminosity from the diffuse intragroup medium
against the fraction of group galaxies with quiescent star
formation. The star formation is determined for galaxies above the mass completeness limit of ZENS, i.e., 
10$^{10}$M$_\odot$, and measured using photometric and spectroscopic measurements and proper
correction for incompleteness~\citep[see][for details]{Cibinel13b}.
In addition, in the right panel of 
Fig.~\ref{f5:fig} the X-ray luminosity is plotted
against the fraction of stellar mass of the group member galaxies that is locked in
the bulge component.  The latter is determined using the bulge-disk
decomposition from both $I$ and $B$ band photometric data, again
using galaxies with mass above $10^{10}$M$_\odot$.  As in previous cases, in both figures data from
XMM-$Newton$ (blue circles) and $Chandra$ (red pentagons) with
signal-to-noise better than two are shown, together with upper limits
for the respective instruments.

The $L_X$ vs quenched fraction could contain information about the
role of the DIM in quenching the star formation in group galaxies,
while conversely the spheroid component, which correlates with black
hole mass, could reveal impact of AGN activity on the DIM.  However,
we see no apparent correlation in either plot using the current data. The small size of the sample and the
numerous upper limits compared to X-ray detections are clearly a limitation in this investigation, which argues for the acquisitions of X-ray data for a substantially larger sample.  While it may sound tempting to
employ a sample with significantly larger range in X-ray luminosity, it would in fact be much more useful to
increase the sample size at fixed group halo mass, so to disentangle additional effects associated with group halo mass. A larger sample is also essential to separately  investigate trends for central and satellite galaxy samples, and for satellites as a function of group-centric distance -- both recognised  key environmental parameters \citet{Weinmann06,Weinmann08,Carollo13}.
 
\section{Conclusions}\label{conc:se}

We have conducted a pilot program of X-ray observations  with the XMM-$Newton$ and $Chandra$ telescopes of a small subset of optically-selected groups belonging to the sample of 141 groups of the Zurich Environmental Study (ZENS). Observations with
XMM-$Newton$ were carried out for 8 groups during three observing
cycles. With the addition of archival data for 4 groups, they amount
to a total of 13 group observations. About one third of the data was
lost due to flares.  X-ray emission from the DIM was successfully
detected for 6 groups, in addition to a marginal 1.8 sigma detection.
The $Chandra$ data were collected during Cycle 11.  A total of 13 ZENS
groups were observed and diffuse X-ray emission from the DIM was 
successfully detected above 2 sigmas for 4 groups.

The target groups have been selected to be in a narrow mass scale
about $10^{13} M_\odot$, and at redshift $z=0.05-0.0585$.  The
detections reveal a typical X-ray flux of in the energy band 0.5-2keV
of $\le 10^{-14}$ erg cm$^{-1}$ s$^{-1}$, which is one order of magnitude
fainter than typical ROSAT groups (RASS). However, for many groups we 
were able to obtain only upper limits, indicating that despite
uncertainties in the mass determination of the groups, our detections
are likely probing the upper envelope of the X-ray emitting, optically
selected ZENS groups.

The main results of this exploratory analysis can be summarized as follows:

\begin{itemize}

\item Small groups such as those probed by ZENS survey are
  characterized by large diversity properties and dynamical
  conditions.  The X-ray data, in particular maps of the DIM, reveal
  features that would be missed by the analysis of the optical data
  alone.  They thus provide important complementary information to
  optical data, which is necessary particularly for a robust
  classification of the dynamical conditions of the group.

\item Our investigation  of the X-ray luminosity versus group mass indicates that
  our optically selected ZENS groups may be under-luminous with respect
  to the prediction from scaling relations characterizing X-ray
  selected groups~\cite{Leauthaud10}, as they crowd almost exclusively
  the space to the left of the $M_X$ vs $L_X$ line corresponding to the said
  scaling relation, the resulting asymmetry being even stronger than
  reported in~\citet{Connelly12}.

\item The mass determination based on the total optical luminosity of
  the groups is in better agreement with the mass determination based
  on X-ray emission than the dynamical mass estimates. 
  The optical masses, however, remain always larger than the X-ray masses estimates,
  indicating the existence of a residual relative bias between the two
  mass proxies. There is tentative evidence that this discrepancy is 
reduced when the group mass determination is based on the luminosity 
of the bulge component of the member galaxies, rather than the total 
luminosity~\cite[consistent with][at larger mass scales]{Andreon12}.
Note that the dynamical mass estimates could also be improved with refined determinations
of the group member galaxies velocity dispersion as proposed recently in~\cite{Erfanianfar13}.

\item The group stellar mass fraction, obtained from the ratio of the
  groups total stellar mass and the total group mass determined from
  the X-ray emission, has a median value of 0.011. This is consistent
  with recent work of~\citet{Connelly12}. The contribution to the
  stellar mass fraction from the most massive galaxies ranges between
  30 to 50 \%.

\end{itemize}

It is clear that the   small  number of diffuse X-ray measurements for    $\sim10^{13}M_\odot$ groups limits   at the moment our ability to  robustly probe the effects of the interplay between the DIM and the  member galaxies at this mass scale that is likely very relevant for environmentally-driven galactic evolution. This pilot program  demonstrates however the potential return and importance of conducting a similar   X-ray study  on a group sample which is large enough to enable a statistically robust investigation of this crucial and yet unexplored issue in galaxy evolution.

\acknowledgments

\clearpage

\begin{deluxetable}{lcccccccccc}
\tabletypesize{\scriptsize}
\tablecaption{XMM-$Newton$'s log of ZENS Galaxy Group Observations\label{xmm_sample}}
\tablewidth{0pt}
\tablehead{
\colhead{Target}&
\colhead{OBS ID}&
\colhead{RA} &
\colhead{DEC}&
\colhead{Redshift}&
\colhead{$\#$ of}&
\colhead{M$_{Group}$} &
\colhead{Cicle}&
\colhead{Exp}&
\colhead{Note\tablenotemark{a}}\\
& & 
\colhead{(J2000)}&
\colhead{(J2000)}&
&
\colhead{galaxies}&
\colhead{$(10^{13}$M$_{\odot})$}&
&
(ksec)&
}
\startdata
2PIGG\_s1520  & 0655300101 & 00:02:01.79 & -34:52:55.5  & 0.05434  &  9 & 1.55 & AO-9  & 33.0 &  \\
2PIGG\_s1571  & 0655300301 & 02:37:04.33 & -25:23:34.3  & 0.0568   & 10 & 1.52 & AO-9  & 38.8 &  \\
2PIGG\_s1783  & 0655300601 & 22:17:26.33 & -36:59:48.1  & 0.05833  &  8 & 4.90 & AO-9  & 39.4 &  \\
2PIGG\_s1614  & 0674480301 & 22:25:15.88 & -25 23 15.4  & 0.05676  & 18 & 4.98 & AO-10 & 15.5 & F \\
2PIGG\_s1471  & 0674480901 & 23:45:01.81 & -26:37:26.8  & 0.05276  & 15 & 3.91 & AO-10 & 13.0 &  \\
2PIGG\_n1466  & 0674480401 & 14:04:01.63 & -01:40:06.9  & 0.05292  & 17 & 4.980 & AO-10 & 15.0 & F \\
2PIGG\_n1572  & 0674480701 & 14:25:33.40 & -01:30:00.4  & 0.05501  & 19 & 4.72 & AO-10 & 15.0 & \\
2PIGG\_s1799  & 0693741001 & 01:14:34.43 & -33:56:09.8  & 0.05819  & 13 & 3.35  & AO-11 & 25.0 & F \\
2PIGG\_n1606  & 0693741601 & 10:38:36.50  & +01:46:01.2 & 0.056  & 7  & 1.55  & AO-11  & 25.7  &  \\
2PIGG\_n1714 & 0150870401 & 10:36:06.57 & -04:02:11.8  & 0.0576   &  7 & 2.00 & AO-2  & 32.5 & A \\
2PIGG\_n1377 & 0207060301 & 11:32:42.10 & -03:50:20.0  & 0.05154  & 23 & 7.51 & AO-3  & 27.7 & A,F \\
2PIGG\_n1330 & 0305800101 & 10:27:36.72 & -03:03:58.8  & 0.05044  &  5 & 1.02 & AO-4  & 24.9 & A \\
2PIGG\_-s1783 & 0550460801 & 22:17:26.33 & -36:59:48.1  & 0.05833  &  8 & 4.90 & AO-7  & 28.0 & A \\
\enddata
\tablenotetext{a}{F=flared, A=Archival Data}
\end{deluxetable}

\begin{deluxetable}{lccccccccccc}
\tabletypesize{\scriptsize}
\tablecaption{XMM-$Newton$ Data on Diffuse Intragroup Emission\label{xmm_extended_sources}}
\tablewidth{0pt}
\tablehead{
\colhead{Target}&
\colhead{Total}&\colhead{Bkdg}&\colhead{Point}&\colhead{Rad}&
\colhead{L$_x^a$}&\colhead{M$_{200}$}&\colhead{R$_{200}$}&\colhead{T$_x$}&\colhead{Flux$^b$}&\colhead{S/N}
\\
&\colhead{cnts}&\colhead{cnts}&\colhead{src}&\colhead{($\arcmin$)}
& \colhead{($10^{41}$erg/s)} 
& \colhead{($10^{13}$M$_\sun$)} 
&\colhead{($\deg$)}&\colhead{(keV)}
&\colhead{($10^{-14}$erg/s/cm$^2$)} &
}
\startdata
2PIGG\_s1520&4707$\pm$69& 3576 & 608 & 3.3 &3.38$\pm$0.50  & 1.29$\pm$0.11 & 0.126  & 0.44$\pm$0.02 &   (3.10$\pm$0.46) &  6.77 \\
2PIGG\_s1571&16938$\pm$130& 8534 & 1630 & 4.9 &31.9$\pm$0.68 & 4.96$\pm$0.06 & 0.189 & 0.90$\pm$0.01 &  (27.88$\pm$0.59) & 46.87 \\
2PIGG\_s1783&8564$\pm$92& 5137 & 2957 & 4.0 & 2.28$\pm$0.45  & 1.02$\pm$0.12 & 0.109  & 0.40$\pm$0.02 &   (1.75$\pm$0.35) &  5.04 \\
2PIGG\_s1614&2426$\pm$49& 2294 & 50 & 5.0 & 3.10$\pm$1.70  & 1.23$\pm$0.37 & 0.119  & 0.43$\pm$0.05 &   (2.57$\pm$1.41) &  1.82\tablenotemark{c} \\   
2PIGG\_s1471&3235$\pm$57& 2518 & 607 & 5.0 & 1.75$\pm$0.82  & 0.87$\pm$0.23 & 0.114  & 0.38$\pm$0.04 &   (1.63$\pm$0.76) &  2.13\tablenotemark{c} \\   
2PIGG\_n1572&3998$\pm$63& 3650 & 377 & 5.0 &$<$1.69$\pm$0.85  &  $<$0.85$\pm$0.24 & $<$0.109  & $<$0.37$\pm$0.04 &   $<$(1.43$\pm$0.72) & -0.93\tablenotemark{d} \\   
2PIGG\_s1799 &4367$\pm$66& 4014 & 78 & 4.0 & $<$7.34$\pm$1.77  & $<$2.05$\pm$0.28 & $<$0.138  & $<$0.55$\pm$0.04 &   $<$(6.08$\pm$1.47) &  4.13\tablenotemark{e} \\  
2PIGG\_n1606  & 1558$\pm$39& 910  & 411 & 2.4 & 2.34$\pm$0.40  & 1.03$\pm$0.10 & 0.113  & 0.40$\pm$0.02 &   (1.95$\pm$0.34) &  5.80 \\
2PIGG\_n1377 & 328$\pm$18& 232 & 14 & 1.5 & 6.26$\pm$1.44  & 1.87$\pm$0.25 & 0.150  & 0.52$\pm$0.03 &   (6.64$\pm$1.53) &  4.35  \\
\enddata
\tablenotetext{a}{0.1-2.4keV}
\tablenotetext{b}{0.5-2keV}
\tablenotetext{c}{A 2 sigma detection}
\tablenotetext{d}{A 2 sigma-limit is calculated}
\tablenotetext{e}{Upper limit}
\end{deluxetable}

\begin{deluxetable}{lcccccccccc}
\tabletypesize{\scriptsize}
\tablecaption{$Chandra$'s log of ZENS Galaxy Group Observations\label{chandra_sample}}
\tablewidth{0pt}
\tablehead{
\colhead{Target}&
\colhead{OBS ID}&
\colhead{RA} &
\colhead{DEC}&
\colhead{Redshift}&
\colhead{$\#$ of}&
\colhead{M$_{Group}$} &
\colhead{Cicle}&
\colhead{Exp}&
\colhead{Note}\\
& & 
\colhead{(J2000)}&
\colhead{(J2000)}&
&
\colhead{galaxies}&
\colhead{$(10^{13}$M$_{\odot})$}&
&
(ksec)&
}
\startdata
2PIGG\_s1571 &  11613             & 02:37:04.33&-25:23:34.3&0.0568&10 &1.52& 11 & 10.06 & \\
2PIGG\_n1610 & 11617-11620 & 09:53:38.23&-05:08:21.4&0.0562&10 &1.45&  11 & 10.03 & \\
2PIGG\_n1702 & 11621-11624 & 09:54:30.67&-04:06:03.3&0.0574& 9 &2.26 &  11  & 9.88 & \\
2PIGG\_n1347 & 11625, 11627 & 09:59:44.62&-05:16:52.6&0.0521&10 &2.90 & 11  & 10.29 & \\
2PIGG\_n1480 & 11629, 11631 &  10:15:31.91&-05:37:06.9&0.0537& 13 &2.26 & 11  & 9.97 & \\
2PIGG\_n1320 & 11633-11636 & 10:17:55.04&-01:22:53.4&0.0508&10 &3.00 & 11  & 10.05 & \\
2PIGG\_n1441 &  11637, 11639 & 11:18:10.68& -04:27:36.1&0.0531&15 &3.41 & 11  & 9.97 & \\
2PIGG\_n1381 & 11641, 11643, 11644 &  14:28:12.53&-02:31:12.4&0.0522&10 &1.22 & 11  & 10.16 & \\
2PIGG\_n1598 & 11645 &  14:35:54.08&-01:16:42.7&0.0560& 9 &2.67 & 11   & 9.79 & \\
2PIGG\_n1746 & 11649, 11652 & 14:40:20.07&-03:45:56.2 &0.0585& 9 &1.65 & 11  & 10.18 & \\
2PIGG\_s1752 & 11653, 11655 & 22:21:10.68&-26:00:24.6&0.0577& 11 &5.60 & 11  & 10.35 & \\
2PIGG\_s1671 & 11657-11660 & 22:24:00.14& -30:00:17.9&0.0567& 10 &2.83 &  11  & 10.58& \\
\enddata
\end{deluxetable}

\begin{deluxetable}{lllllllllll}
\tabletypesize{\scriptsize}
\tablecaption{$Chandra$ Data on Diffuse Intragroup Emission\label{chandra_extended_sources}}
\tablewidth{0pt}
\tablehead{
\colhead{Target}&
\colhead{Total}&\colhead{Bkdg}&\colhead{Point}&\colhead{Rad}&
\colhead{L$_x^a$}&\colhead{M$_{200}$}&\colhead{R$_{200}$}&\colhead{T$_x$}&\colhead{Flux$^b$}&\colhead{S/N}
\\
&\colhead{cnts}&\colhead{cnts}&\colhead{src}&\colhead{($\arcmin$)}
& \colhead{($10^{41}$erg/s)} 
& \colhead{($10^{13}$M$_\sun$)} 
&\colhead{($\deg$)}&\colhead{(keV)}
&\colhead{($10^{-14}$erg/s/cm$^2$)} &
}
\startdata
2PIGG\_s1571&616$\pm$24.8&332&17&3&34.33$\pm$3.19&5.19$\pm$0.28&0.19&0.93$\pm0.03$&30.0$\pm$2.79&10.76\\
2PIGG\_n1610&185$\pm$13.6&179&28.9&2&$<$3.72$\pm$1.86&$<$1.37$\pm$0.24&&&$<$3.20$\pm$1.60&-1.68\\
2PIGG\_n1702&86.9$\pm$9.3&65&0&1.5&2.84$\pm$1.26&1.16$\pm$0.24&0.12&0.42$\pm$0.04&2.29$\pm$1.02&2.26\\
2PIGG\_n1347&194$\pm$13.9&133&34.6&2&$<$3.26$\pm$1.63&$<$1.27$\pm$0.35&&&$<$3.24$\pm$1.62&1.90\\
2PIGG\_n1480&157$\pm$12.5&151&7.6&2&$<$3.07$\pm$1.54&$<$1.22$\pm$0.34&&&$<$2.86$\pm$1.43&-0.13\\
2PIGG\_n1320&133$\pm$11.5&68&14.7&1.4&5.39$\pm$1.24&1.71$\pm$0.23&0.15&0.50$\pm$0.03&5.86$\pm$1.34&4.36\\
2PIGG\_n1441&199$\pm$14.1&155&20.1&2&$<$3.46$\pm$1.73&$<$1.31$\pm$0.36&&&$<$3.32$\pm$1.66&1.70\\
2PIGG\_n1381&208$\pm$14.4&159&37.8&2&$<$3.18$\pm$1.59&$<$1.25$\pm$0.34&&&$<$3.16$\pm$1.58&0.78\\
2PIGG\_n1598&184$\pm$13.6&141&23.7&2&$<$3.69$\pm$1.84&$<$1.36$\pm$0.37&&&$<$3.19$\pm$1.60&1.42\\
2PIGG\_n1746&62$\pm$7.9&31&0&0.9&5.08$\pm$1.29&1.65$\pm$0.24&0.13&0.49$\pm$0.03&4.09$\pm$1.04&3.92\\
2PIGG\_s1752&374$\pm$19.3&372&16&3&$<$5.16$\pm$2.58&$<$1.66$\pm$0.46&&&$<$4.27$\pm$2.14&-0.72\\
2PIGG\_s1671&438$\pm$20.9&391&27&3&$<$5.36$\pm$2.68&$<$1.70$\pm$0.47&&&$<$4.62$\pm$2.31&0.96\\
\enddata
\tablenotetext{a}{0.1-2.4keV}
\tablenotetext{b}{0.5-2keV}
\end{deluxetable}

\begin{deluxetable}{lcccccc}
\tabletypesize{\scriptsize}
\tablecaption{Optical vs X-ray Centers\label{matches:tab}}
\tablewidth{0pt}
\tablehead{
\colhead{Group}&
\colhead{RA} &
\colhead{DEC}&
\colhead{$\Delta\alpha_{\rm Opt-Xray}$}&
\colhead{$\Delta\alpha_{\rm Opt-CG}$}&
\colhead{$\Delta\alpha_{\rm CG-Xray}$}&
\colhead{State\tablenotemark{a}}
\\
& 
\colhead{(J2000)}&
\colhead{(J2000)}&
\colhead{('')} &
\colhead{('')} &
\colhead{('')} &
}
\startdata
2PIGG\_s1520 & 00:02:01.79 & -34:52:55.5 & 3.3 & 0.23 & 3.3 &  U \\
2PIGG\_s1571 & 02:37:04.33 & -25:23:34.3 & 7.4 & 0.24 & 7.3 &  R \\
2PIGG\_n1377 & 11:32:42.10 & -03:50:20.0 & 60  & 718  & 695 &  U \\
2PIGG\_n1320 & 10:17:55.04 & -01:22:53.4 & 1.8 & 312  & 310 &  R \\
2PIGG\_n1746 & 14:40:20.07 & -03:45:56.2 & 36  & 38   & 10 &  R \\
2PIGG\_n1606 & 10:38:36.50 & 01:46:01.2  &280 & 268 & 36 & R
\enddata
\tablenotetext{a}{R=relaxed, U=unrelaxed}
\end{deluxetable}

\clearpage

\end{document}